\documentclass[times,twocolumn]{aastex631}
\usepackage{CJK}
\usepackage{amsmath, amssymb, bm}
\usepackage{mathrsfs}
\usepackage[nodayofweek]{datetime}

\newcommand{\uat}[2]{\href{http://astrothesaurus.org/uat/#1}{#2  (#1)}}
\newcommand{\hi}{\rm H\,{\textsc{\romannumeral 1}}}
\newcommand{\nhi}{{N_{\rm H\,{\textsc{\romannumeral 1}}}}}

\newcommand{\shi}{{S_{\rm H\,{\textsc{\romannumeral 1}}}}}

\received{November 19, 2022}
\revised{May 30, 2023}
\accepted{June 1, 2023}
\published{July 25, 2023}

\makeatletter
\patchcmd{\NAT@citex}
  {\@citea\NAT@hyper@{%
     \NAT@nmfmt{\NAT@nm}%
     \hyper@natlinkbreak{\NAT@aysep\NAT@spacechar}{\@citeb\@extra@b@citeb}%
     \NAT@date}}
  {\@citea\NAT@nmfmt{\NAT@nm}%
   \NAT@aysep\NAT@spacechar\NAT@hyper@{\NAT@date}}{}{}

\patchcmd{\NAT@citex}
  {\@citea\NAT@hyper@{%
     \NAT@nmfmt{\NAT@nm}%
     \hyper@natlinkbreak{\NAT@spacechar\NAT@@open\if*#1*\else#1\NAT@spacechar\fi}%
       {\@citeb\@extra@b@citeb}%
     \NAT@date}}
  {\@citea\NAT@nmfmt{\NAT@nm}%
   \NAT@spacechar\NAT@@open\if*#1*\else#1\NAT@spacechar\fi\NAT@hyper@{\NAT@date}}
  {}{}
\makeatother

\shortauthors{Yu et al.}

\begin{document}
\begin{CJK*}{UTF8}{gbsn}
\title{\hi\ Absorption in Low-power Radio AGNs Detected by FAST}

\correspondingauthor{Taotao Fang}

\author[0000-0003-3230-3981]{Qingzheng Yu (余清正)}
\affiliation{Department of Astronomy, Xiamen University, Xiamen, Fujian 361005, People's Republic of China; \url{fangt@xmu.edu.cn, yuqingzheng@stu.xmu.edu.cn}}

\author[0000-0002-2853-3808]{Taotao Fang (方陶陶)}
\affiliation{Department of Astronomy, Xiamen University, Xiamen, Fujian 361005, People's Republic of China; \url{fangt@xmu.edu.cn, yuqingzheng@stu.xmu.edu.cn}}

\author[0000-0003-4874-0369]{Junfeng Wang (王俊峰)}
\affiliation{Department of Astronomy, Xiamen University, Xiamen, Fujian 361005, People's Republic of China; \url{fangt@xmu.edu.cn, yuqingzheng@stu.xmu.edu.cn}}

\author[0000-0001-7349-4695]{Jianfeng Wu (武剑锋)}
\affiliation{Department of Astronomy, Xiamen University, Xiamen, Fujian 361005, People's Republic of China; \url{fangt@xmu.edu.cn, yuqingzheng@stu.xmu.edu.cn}}



\begin{abstract}

\noindent We report the discovery of three \hi\ absorbers toward low-power radio active galactic nuclei (AGNs) in a pilot \hi\ absorption survey with the Five-hundred-meter Aperture Spherical radio Telescope (FAST). Compared to past studies, FAST observations have explored lower radio powers by $\sim$0.4 dex and detected these weakest absorbers at given redshifts. By comparing the gas properties and kinematics of sources along radio powers, we aim to explore the interplay between AGN and the surrounding interstellar medium (ISM). Compared to brighter sources at similar redshifts, our observations suggest a slightly lower detection rate of \hi\ absorption lines ($\sim$$ 11.5\%$) in low-power radio AGNs with $\text{log}(P_{\text{1.4 GHz}}/\text{W Hz}^{-1})=21.8-23.7$. The low-power sources with $\text{log}(P_{\text{1.4 GHz}}/\text{W Hz}^{-1})<23$ have a lower detection rate of $\sim$$ 6.7\%$. Due to the incompleteness of the sample, these detection rates may represent the lower limits. The selection of more extended sources and dilution by \hi\ emission at lower redshifts may contribute to the lower detection rate of \hi\ absorption lines. These detected absorbers present relatively narrow line widths and comparable column densities consistent with previous observations. One absorber has a symmetric profile with a large velocity offset, while the other two show asymmetric profiles that can be decomposed into multiple components, suggesting various possibilities of gas origins and kinematics. These \hi\ absorbers may have connections with rotating disks, gas outflows, galactic gas clouds, gas fueling of the AGN, and jet-ISM interactions, which will be further investigated with the upcoming systematic survey and spatially resolved observations.

\end{abstract}


\section{Introduction} \label{sec:intro}

The associated \hi\ 21 cm absorption provides an important method to study the physical conditions of active galactic nuclei (AGNs) and their interplay with the interstellar medium (ISM) of host galaxies \cite[for a review see][]{Morganti2018}. Previous observations of \hi\ absorption lines in radio AGNs show that the \hi\ gas can be traced in regularly rotating large-scale disks or circumnuclear disks \citep{Gallimore1999,Struve2010,Morganti2011}. Kinematics of the \hi\ absorption lines also have been used to reveal infalling clouds correlated to the feeding of the supermassive black hole (SMBH) \citep{Morganti2009,Maccagni2014,Tremblay2016}, as well as outflows or galactic gas clouds indicating the interactions between jets and surrounding ISM \citep{Conway1999,Morganti2013,Aditya2018,Schulz2021}. Observations of \hi\ absorption lines in different types of AGNs can provide information on the evolution of AGNs and host galaxies, e.g., some recent studies suggest the \hi\ absorption is tightly correlated to galaxy mergers \citep{Dutta2018,Dutta2019,Dutta2022}. 

Although \hi\ absorption has been studied for decades and more than 100 \hi\ absorbers have been detected at $z<1$ \citep{Morganti2018}, our understanding is still limited on the \hi\ absorption and related evolution processes in various classes of AGN, especially the faint low-power radio AGNs. Due to the sensitivity limits, most \hi\ 21 cm absorption surveys are limited to bright sources with $S_{\rm 1.4 GHz}>$ 50 mJy \citep{Gallimore1999,Ger2015,Aditya2019}. A recent survey carried out with the Westerbork Synthesis Radio Telescope (WSRT) has expanded the sample to weaker sources with $S_{\rm 1.4 GHz}>$ 30 mJy \citep{Maccagni2017} in the redshift range $0.02<z<0.25$, which increases detections of \hi\ absorption in low-power radio AGNs ($\text{log}(P_{\text{1.4 GHz}}/\text{W Hz}^{-1})<24$). This survey has found a detection rate of $\sim$27\% in the entire sample, and the detection rate remains similar across the range of radio powers. Their results indicate AGNs at the low-power range mostly show narrow absorption features, which traced gas in rotating disks, while the broad \hi\ absorption lines (e.g., fast \hi\ outflows) are only found in high-power radio AGNs or mergers. For the low-power radio AGNs, the number of observed sources with $\text{log}(P_{\text{1.4 GHz}}/\text{W Hz}^{-1})<23$ is smaller than that of brighter sources with $\text{log}(P_{\text{1.4 GHz}}/\text{W Hz}^{-1})>25$ by a factor of $\sim$4 \citep{Maccagni2017,Murthy2021}. It is unclear whether these previous results reveal the nature of \hi\ absorption in low-power sources or have a bias due to the flux-limited sample. Furthermore, larger samples and more observations of low-power radio sources are needed to help understand what drives the observational differences of the \hi\ absorption properties between low-power and high-power sources. In particular, comparisons among sources of all radio powers can provide critical information to study the interplay between the radio activity of AGN and the surrounding ISM. Thus, a systematic search of \hi\ absorption lines in faint AGNs will help understand the possible variations of detection rate, gas properties, and kinematics along radio powers. In addition, further study with the faint AGN sample will help explore the connection between the radio power of the AGN and \hi\ outflows/fueling. On the other hand, detailed observations with high spatial resolution can provide information on the origins and impacts of \hi\ absorbing gas in low-power AGNs, which may help in investigating various underlying physical processes of the \hi\ absorptions and AGN activity between low-power and high-power sources.

Constructed as the most sensitive single-dish radio telescope, the Five-hundred-meter Aperture Spherical radio Telescope \citep[FAST;][]{Nan2006,Nan2011,Jiang2020} has made significant progress in probing the faintest \hi\ gas in our Milky
Way \citep{Hong2022} and the nearby universe \citep{Zhu2021,Xu2022}. FAST has also successfully carried out a blind search and targeted observations of \hi\ 21 cm absorption lines \citep{Zhang2021,Hu2023,Jing2023}. Recently, \cite{Yu2022} discovered a new \hi\ absorber in an interacting galaxy pair observed by FAST. The newly detected \hi\ absorption line has a very faint radio continuum as background \citep{Yu2022}, which shows the ability of FAST to search the \hi\ absorption toward the faint sources. Hence, we are carrying out a systematic survey of \hi\ absorption lines in low-power radio AGNs. In this paper, we present our results from pilot observations of the survey. 

This paper is organized as follows. In Section \ref{sec:obs}, we introduce the selection of the low-power radio AGN sample, observations with FAST, and data reduction. The main results are presented in Section \ref{sec:res}, which includes the detection rate of the \hi\ absorption lines and properties of individual detections. In Section \ref{sec:dis}, we further discuss the results of this study. Finally, we summarize the main results in Section \ref{sec:sum}. Throughout this paper, we adopt $H_0 = 70$ km s$^{-1}$ Mpc$^{-1}$, $\Omega_M = 0.3$, and $\Omega_{\Lambda} = 0.7$.

\section{Observations and Data Reduction} \label{sec:obs}

The three \hi\ absorbers were newly discovered through our pilot FAST PI program (PT2021\_0067 PI: Q. Yu). To investigate the \hi\ absorption properties of low-power radio AGNs, we selected the targets by cross-matching the NRAO Very Large Array (VLA) Sky Survey (NVSS) catalog \citep{Condon1998} and the Sloan Digital Sky Survey Data Release 16 (SDSS DR16) catalog of galaxies \citep{Ahumada2020}. We selected all sources that have radio flux of 10 mJy $<S_{\rm 1.4 GHz}<$ 30 mJy to expand the sample of radio AGNs presented in \cite{Maccagni2017}. We restricted the redshift of sources to $z<0.1$ to avoid strong radio frequency interference (RFI). We crossmatched the selected sources with the second data release of HI-MaNGA \citep{Stark2021} and Arecibo Legacy Fast Arecibo L-Band Feed Array 
Survey (ALFALFA) catalogs \citep{Haynes2018} to exclude the sources that already have \hi\ 21 cm emission line detections at S/N $>$ 5. With the above constraints, we have compiled a sample of 159 low-power radio AGNs. In our pilot observations, 29 sources in the radio power range of $\text{log }(P_{\text{1.4 GHz}}/\text{W Hz}^{-1})=21.8-23.7$ were observed in 2021 August to 2022 May with the FAST 19-beam receiver using the ON-OFF mode. The OFF-target position was set to 11\farcm5 to avoid contamination of radio continuum sources. With a beam size of $\sim$2\farcm9, the 19-beam receiver was configured with the wide-band spectrometer (Spec(W)) to cover a band of 1.05$-$1.45 GHz with a spectral resolution of $\sim$1.6 $\text{km s}^{-1}$. The integration time varies in the range of 3$-$21 minutes for each source based on the flux of the radio continuum at 1.4 GHz \citep{Condon1998}. Our observations are designed to reach a depth comparable to those in  previous surveys \citep{Ger2015,Maccagni2017}, which are sensitive enough to detect absorption lines with a peak optical depth of $\tau_{\text peak}\sim 0.08$.

Data were reduced with {\tt astropy} \citep{Astropy2013} and {\tt scipy} \citep{SciPy2020}. We first calibrated the spectrum with the injected noise signal to derive the antenna temperature \citep{Jiang2020}. After checking the consistency of polarization XX and YY for each spectrum, we combined the two polarizations. Then the bandpass subtraction was applied with calibrated ON-target and OFF-target spectra. The temperature of each spectrum was converted to flux density based on the gain factor of different beams on the receiver (see Table 5 of \citealt{Jiang2020}). We subtracted the baseline of the spectrum by fitting a sinusoidal plus a polynomial function. RFI was flagged manually during the subtraction process, and we excluded the RFI-contaminated data in later works of spectral line fitting and measurement. The velocity of the final spectrum was Doppler corrected and converted to the barycentric frame. After data reduction, our observations achieved a median rms noise of $\sim$0.63 mJy measured at the velocity resolution of $\sim$1.6 $\text{km s}^{-1}$. For the detections, we have rebinned the spectrum to improve the S/N, and the final rms noise is $\sim$0.31$-$0.40 mJy at the velocity resolution of $\sim$10$-$20 $\text{km s}^{-1}$ (Table \ref{tab:res}). For the non-detections, the median rms noise is 0.19 mJy at the velocity resolution of $\sim$16 $\text{km s}^{-1}$. Although the longest integration time of individual sources is less than a half hour, the average noise level of our observations is lower than previous surveys \citep[e.g.,][]{Ger2015,Maccagni2017} by a factor of 4 at similar spectral resolutions.

\begin{figure}[t!]
\includegraphics[width=0.45\textwidth]{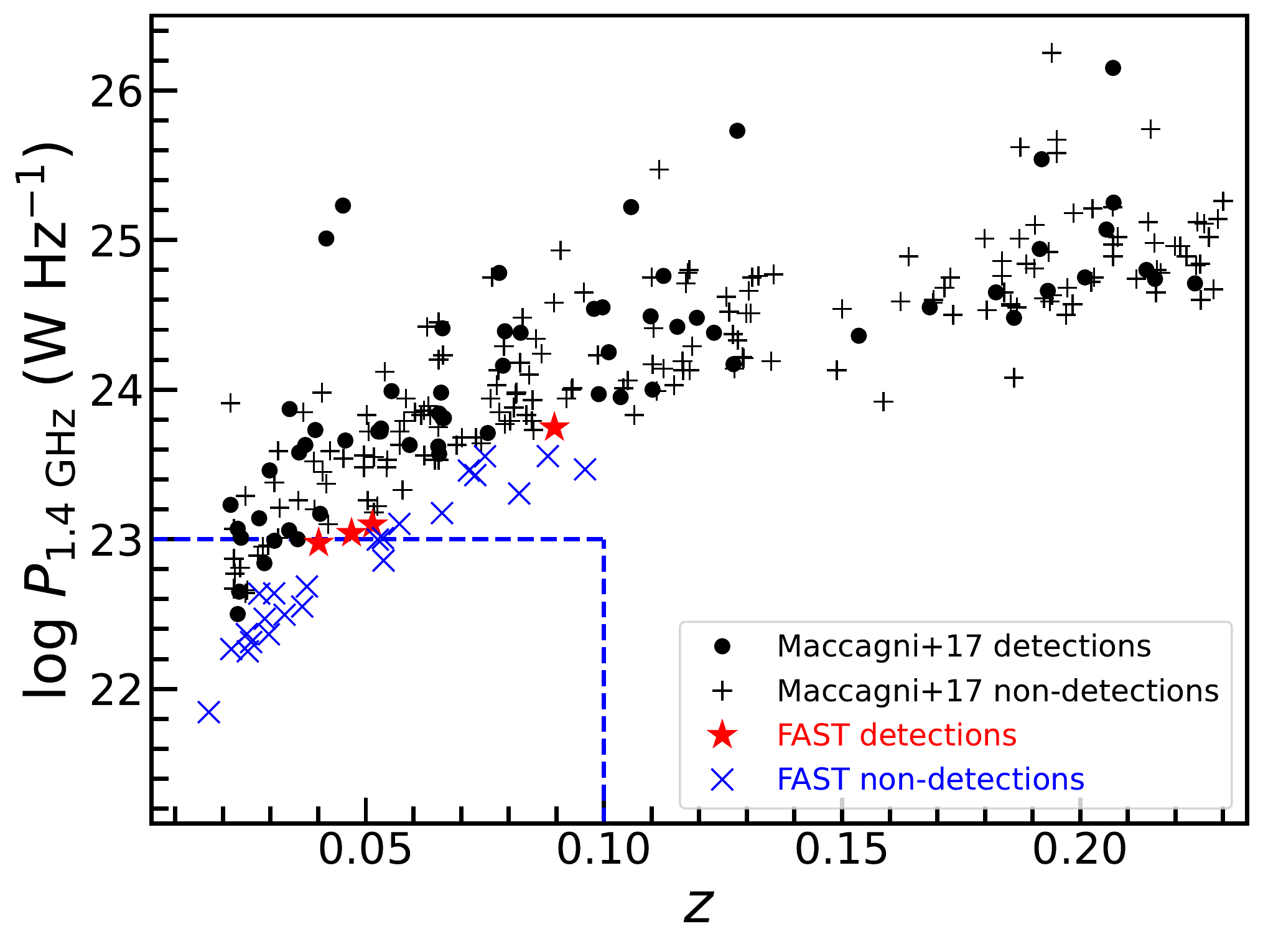}
\caption{Comparison of the radio power and redshift between our observed sources and that of \cite{Maccagni2017}.} 
\label{fig:dist}
\end{figure}

\begin{figure}[t!]
\includegraphics[width=0.45\textwidth]{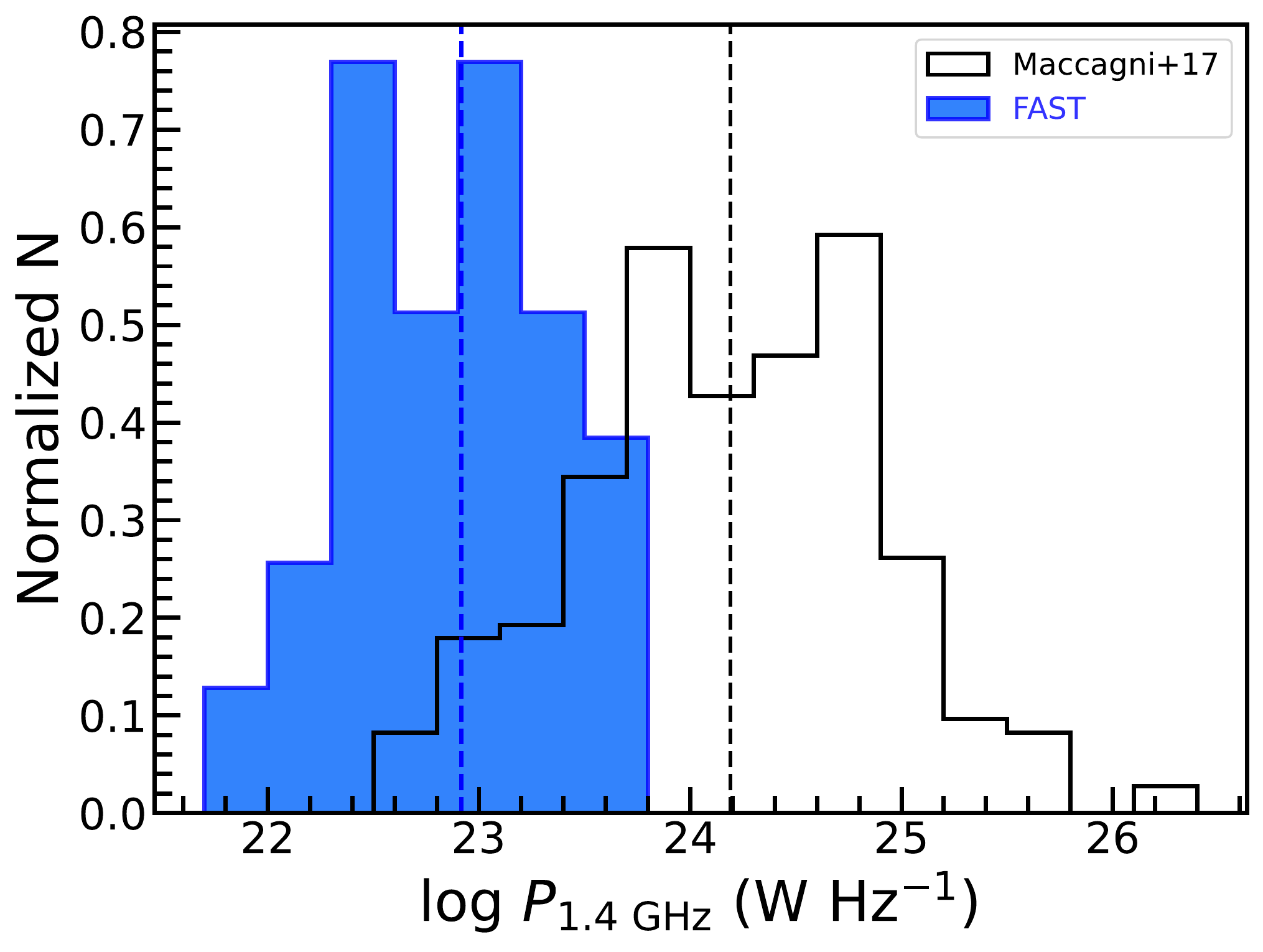}
\caption{Distributions of the radio power of our sample and that of \cite{Maccagni2017}.} 
\label{fig:hist_rp}
\end{figure}

In the following analysis, we also include an \hi\ absorber in an interacting galaxy pair J1558$+$2759 previously discovered by FAST \citep{Yu2022} to investigate the gas properties and origins in Section \ref{sec:dis}. This \hi\ absorption line has a low-power radio AGN as a background continuum source similar to these three absorbers. Therefore, we discuss the \hi\ gas properties using the new FAST observations and previous data in \cite{Yu2022}. However, considering the different observing setups between the new and previous observations, we only include the 29 sources in the pilot observations to discuss the detection rate of \hi\ absorption lines (Section \ref{sec:res}).

\begin{figure}[t!]
\includegraphics[width=0.45\textwidth]{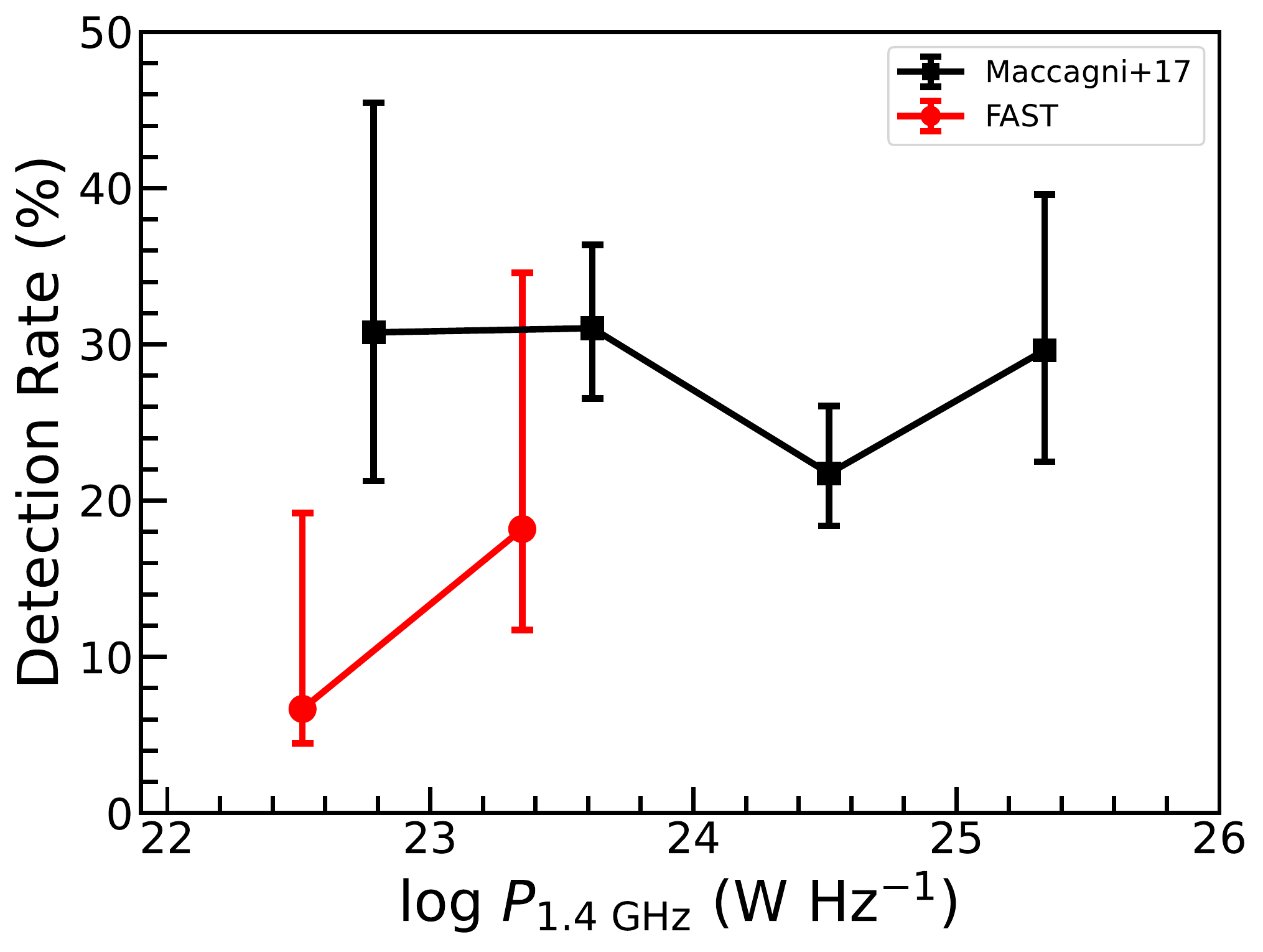}
\caption{Comparison of the detection rate between our sample and that of \cite{Maccagni2017}. The detection rates of each sample are divided into 4 bins based on the radio power. The 4 bins are (1) $\text{log}(P_{\text{1.4 GHz}}/\text{W Hz}^{-1})<23$; (2) $23\leqslant \text{log}(P_{\text{1.4 GHz}}/\text{W Hz}^{-1})<24$; (3) $24\leqslant \text{log}(P_{\text{1.4 GHz}}/\text{W Hz}^{-1})<25$; (4) $\text{log}(P_{\text{1.4 GHz}}/\text{W Hz}^{-1})\geqslant25$.}
\label{fig:detect_rate}
\end{figure}

In Figure \ref{fig:dist}, we plot the distributions of our observed sources and sources detected by \cite{Maccagni2017} in the radio power-redshift space for comparison. Compared to sources in \cite{Maccagni2017}, the radio power of our sample is systematically lower than that of AGNs at similar redshifts, and the detected \hi\ absorbers in our sample are among the weakest of detections at given redshifts. The blue dashed lines marked the region of $\text{log}(P_{\text{1.4 GHz}}/\text{W Hz}^{-1})<23$ and $z<0.1$, which has been less explored in previous surveys. In the radio power range of $\text{log}(P_{\text{1.4 GHz}}/\text{W Hz}^{-1})<23$, our observations have increased the number of observed sources by a factor of 2. As shown in Figure \ref{fig:hist_rp}, we present the histograms of the radio power range for our sample and that of \cite{Maccagni2017}. The radio power of our sample distributes in the range of $\text{log }(P_{\text{1.4 GHz}}/\text{W Hz}^{-1})=21.8-23.7$, and the median value of our sample ($\text{log }(P_{\text{1.4 GHz}}/\text{W Hz}^{-1})=22.9$) is lower than that of \cite{Maccagni2017} ($\text{log }(P_{\text{1.4 GHz}}/\text{W Hz}^{-1})=24.2$). Briefly, through observations of our sample, we aim to explore the \hi\ absorption line in the faint radio AGN population, which has been less investigated in previous observations. Compared with previous observations \citep{Ger2015,Maccagni2017}, our FAST observations on average have explored lower radio powers by $\sim$0.4 dex at given redshifts (Figure \ref{fig:dist}).

\section{Results} \label{sec:res}

The pilot observations presented in this work yield the detection of new \hi\ absorption lines in three galaxies with S/N $>4.5$. The detection of new \hi\ absorbers suggests our observations can reach the sensitivity limit of absorption lines with the peak optical depth of $\tau_{\text peak}\sim 0.04$. Excluding three sources severely contaminated by the RFI, our pilot observations detected \hi\ absorption lines in three out of 26 sources, suggesting a detection rate of \hi\ absorption lines in low-power radio AGNs that is $\sim$$11.5^{+9.2}_{-3.7}\%$ in the radio power range of $\text{log }(P_{\text{1.4 GHz}}/\text{W Hz}^{-1})=21.8-23.7\ $. The detection rate is relatively lower than that of previous surveys \citep{Ger2015,Maccagni2017}, but the difference is not significant ($< 2\sigma$).

To investigate possible variations of the detection rate with the radio power, we plot the detection rate of our sample and that of \cite{Maccagni2017} along the radio power in Figure \ref{fig:detect_rate}. The sources of each sample were divided into 4 bins based on radio powers, and the detection rate was calculated for each radio power bin. Since the sizes of subsamples in each bin are different and some are relatively small, we estimate the 1$\sigma$ error bars based on the binomial statistics of \cite{Cameron2011}. The results of \cite{Maccagni2017} suggest the detection rate is similar along the entire range of radio powers in their survey. In contrast, our sample shows a lower detection rate in the radio power range of $\text{log}(P_{\text{1.4 GHz}}/\text{W Hz}^{-1})<23$. As shown in Figure \ref{fig:detect_rate}, the detection rate of sources with $\text{log}(P_{\text{1.4 GHz}}/\text{W Hz}^{-1})<23$ from \cite{Maccagni2017} is higher than the detection rate of our sources ($\sim$$6.7^{+12.5}_{-2.2}\%$) by a factor of $\sim$4.6. Although the difference between these two subsamples is not significant ($<3\sigma$) considering the error bars, it is worth further investigating with larger samples to confirm whether the low-power radio AGNs with $\text{log}(P_{\text{1.4 GHz}}/\text{W Hz}^{-1})<23$ have a lower detection rate of \hi\ absorption. In the radio power range of $23\leqslant \text{log}(P_{\text{1.4 GHz}}/\text{W Hz}^{-1})<24$, the detection rate of our sample is consistent with that of \cite{Maccagni2017} within the error bars. Our results indicate that the low-power sources with $\text{log}(P_{\text{1.4 GHz}}/\text{W Hz}^{-1})<23$ may have a lower detection rate of \hi\ absorption, and the detection rate increases and remains similar as the radio power elevating.

\begin{figure*}[t]
\includegraphics[width=\textwidth]{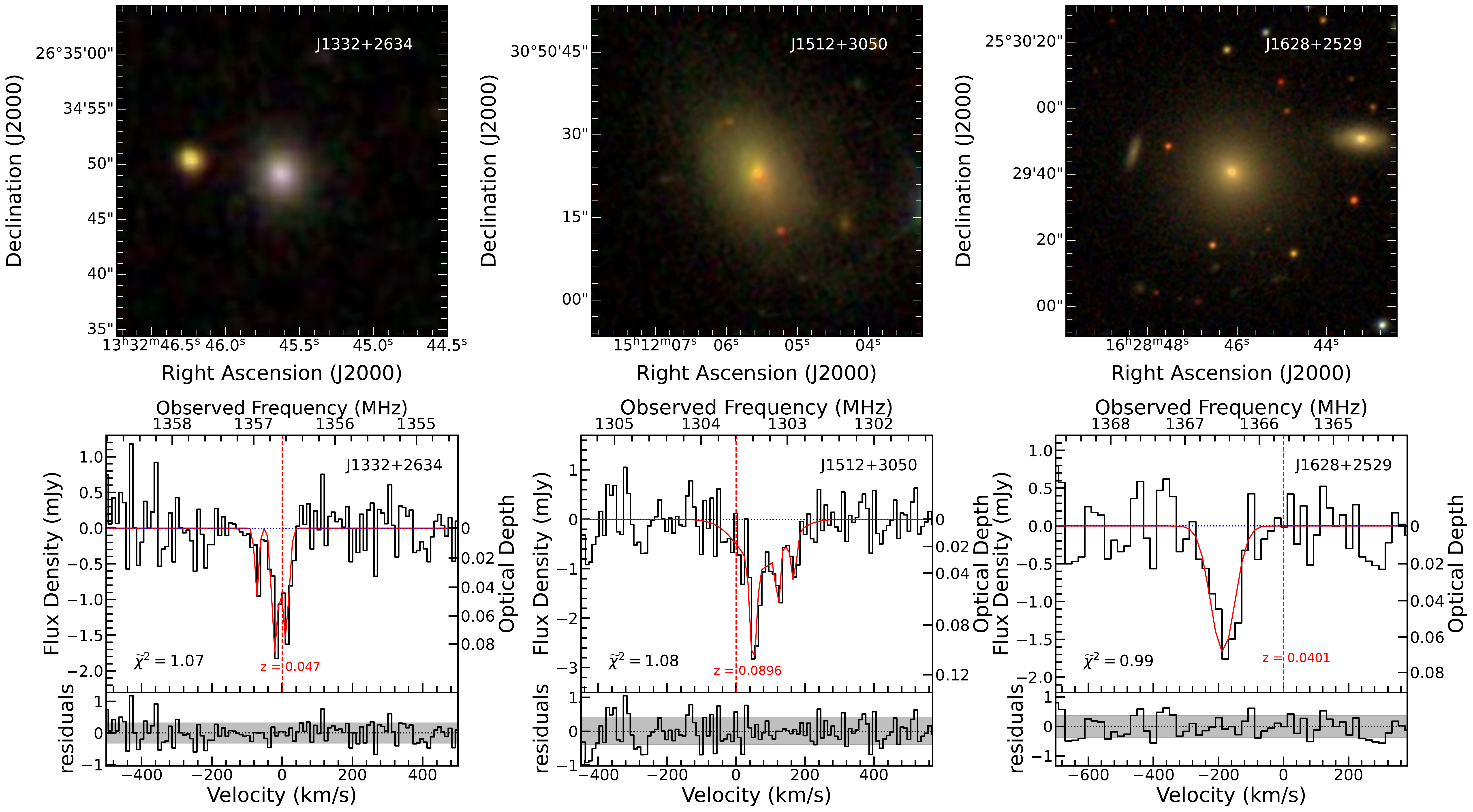}
\caption{The upper panels show SDSS composite images of each source. The FAST beam size ($\sim$2\farcm9) is much larger than the size of these optical images. The lower panels show the corresponding \hi\ spectra of J1332$+$2634,  J1512$+$3050, and J1628$+$2529. The velocity resolution of J1332$+$2634 and J1512$+$3050 is $\sim$$10\ \text{km s}^{-1}$, while the spectrum of J1628$+$2529 is rebinned to $\sim$$20\ \text{km s}^{-1}$. The red lines in each panel show the profiles of Gaussian fits. On the bottom of each panel, the residuals of the fit are plotted with the 1$\sigma$ noise values presented with shaded regions. }
\label{fig:abs_profile}
\end{figure*}

As shown in Figure \ref{fig:abs_profile}, two \hi\ absorbers present multi-peak profiles, and each peak is identified with an S/N $>3$. Therefore, we have performed a multicomponent Gaussian fit to measure the properties of each absorption line. In the multicomponent Gaussian models, we decided the number of components by requiring the reduced $\chi^2$ to be closest to unity. We have checked the Akaike information criterion (AIC) between a single Gaussian fit and a multi-Gaussian fit, and the AIC suggests the multi-Gaussian model can better fit the data. The $\chi^2$, degree of freedom (d.o.f.), and AIC of both single-component and multicomponent Gaussian fits are listed in Table \ref{tab:guass}. The results of multicomponent Gaussian fit revealed the asymmetric shape of the line profiles, indicating the absorbing gas may have multiple velocity components. However, these components still need to be confirmed with higher spectral resolution and S/N. We summarize the measured parameters and the results of multicomponent Gaussian fit in Table \ref{tab:res} and Table \ref{tab:fit}, respectively. The line width $W_{20}$ is measured at 20\% of peak flux density for each Gaussian fit profile, and the line centroids of each absorption system are measured at the center point of $W_{20}$. The column density is calculated with the following equation: 

\begin{equation}
\nhi = 1.823 \times 10^{18}\ {T_s} \int \tau(v) {\rm d} v \text{ cm}^{-2},  
\label{eq:1}
\end{equation}
where ${T_s}$ represents the spin temperature in Kelvin, and $v$ is the Doppler-corrected velocity in kilometer per second. The optical depth $\tau(v)$ is derived by:

\begin{equation}
\tau(v) = - \text{ln}\left(1 + \frac{\shi}{c_f\ S_{\text{1.4 GHz}}} \right) ,  
\label{eq:2}
\end{equation}
where $\shi$ is the line flux of the \hi\ absorption, $S_{\text{1.4 GHz}}$ is the continuum flux at 1.4 GHz, and ${c_f}$ is the covering factor of the gas. In this work, we assume the spin temperature ${T_s}$ = 100 K and the covering factor ${c_f} = 1$ for each source, which has been commonly adopted in previous studies \citep[e.g.,][]{Ger2015,Maccagni2017,Chowdhury2020}.

\begin{deluxetable*}{lcccccccccccrr}\setlength\tabcolsep{2pt}
\tablenum{1}
\tablecaption{Absorption line properties\label{tab:res}}
\tabletypesize{\footnotesize}
\tablewidth{\textwidth}
\tablehead{
\colhead{Source name} & \colhead{R.A.} & \colhead{Decl.} &
\colhead{$t_{\text{ON}}$} & \colhead{$z$} & \colhead{$S_{\text{1.4\ GHz}}$} & \colhead{log($P_{\text{1.4\ GHz}}$)} & \colhead{rms} & \colhead{$\Delta v$} & \colhead{$W_{20}$} & \colhead{$v_{\text{centroid}}$} & \colhead{$\tau_{\text{peak}}$} & \colhead{$\tau_{\text{int}}$} & \colhead{$\nhi$}\\
\colhead{} & \colhead{deg.} & \colhead{deg.} &
\colhead{(min)} & \colhead{} & \colhead{(mJy)} & \colhead{(W $\text{Hz}^{-1}$)} & \colhead{(mJy)} & \colhead{($\text{km s}^{-1}$)}  & \colhead{($\text{km s}^{-1}$)} & \colhead{($\text{km s}^{-1}$)} & \colhead{} & \colhead{($\text{km s}^{-1}$)} & \colhead{($10^{20} \text{cm}^{-2}$)}
}
\decimalcolnumbers
\startdata
J1332$+$2634 & 203.19009 & 26.58038 & 5 & 0.04701 & 21.1 & 23.04 & 0.31 & 10 & 106.9 & $-$26.8 & 0.091 $\pm$ 0.017 & 4.91 $\pm$ 0.67 & 8.95 $\pm$ 1.22 \\
J1512$+$3050 & 228.02324 & 30.83984 & 6.5 & 0.08963 & 27.8 & 23.74 & 0.40 & 10 & 172.7 & 93.0 & 0.107 $\pm$ 0.019 & 8.81 $\pm$ 0.81 & 16.06 $\pm$ 1.48 \\
J1558$+$2759\tablenotemark{a} & 239.75275 & 27.98537 & 13 & 0.05140 & 20.0 & 23.10 & 0.44 & 1.6 & 72.3 & 125.3 &  0.490 $\pm$ 0.050 & 14.50 $\pm$ 0.32 & 47.30 $\pm$ 1.01\\ 
J1628$+$2529 & 247.19221 & 25.49472 & 3.3 & 0.04010 & 25.2 & 22.98 & 0.38 & 20 & 131.6 & $-$195.9 & 0.072 $\pm$ 0.017 & 6.46 $\pm$ 0.97 & 11.78 $\pm$ 1.77
\enddata
\tablecomments{The columns are (1) the source name of the AGN; (2) R.A. in degrees; (3) Decl. in degrees; (4) integration time; (5) redshift from SDSS spectroscopy \citep{Abolfathi2018}; (6) radio continuum flux at 1.4 GHz from NVSS \citep{Condon1998}; (7) radio power at 1.4 GHz; (8) 1$\sigma$ noise measured at re-binned velocity resolution; (9) re-binned velocity resolution; (10) line width measured at 20\% of the peak flux; (11) line centroid of the absorption; (12) peak optical depth; (13) integrated optical depth; (14) estimated \hi\ column density assuming $T_s=100$ K, $c_f=1$.}
\tablenotetext{a}{The detailed data and measurements of this source are adopted from \cite{Yu2022}.}
\end{deluxetable*}

\begin{deluxetable}{lcccccc}\setlength\tabcolsep{1.5pt}
\tablenum{2}
\tablecaption{Comparison of single and multiple Gaussian fits for the multi-peak \hi\ absorbers \label{tab:guass}}
\tabletypesize{\footnotesize}
\tablewidth{\textwidth}
\tablehead{
\colhead{Source name} & \colhead{$\chi^2_{\rm single}$} & \colhead{$\rm d.o.f._{single}$} & \colhead{$\rm AIC_{single}$} & \colhead{$\chi^2_{\rm multi}$} & \colhead{$\rm d.o.f._{multi}$} & \colhead{$\rm AIC_{multi}$}
}
\decimalcolnumbers
\startdata
J1332$+$2634 & 270.3 & 237 & 34.5 & 247.0 & 231 & 24.9  \\
J1512$+$3050 & 136.0 & 100 & 34.6 & 98.7 & 91 & 19.6 
\enddata
\tablecomments{The columns are (1) the source name of the AGN; (2) $\chi^2$ of the single-component Gaussian fit; (3) the degree of freedom (d.o.f.) of the single-component Gaussian fit; (4) Akaike information criterion (AIC) of the single-component Gaussian fit; (5) $\chi^2$ of the multi-component Gaussian fit; (6) d.o.f. of the multi-component Gaussian fit; (7) AIC of the multi-component Gaussian fit.}
\end{deluxetable}

\subsection{J1332\sl{+}2634} \label{subsec:J1332+2634}

The source J1332$+$2634 is an isolated galaxy with a spectroscopic redshift of $z=0.04701\pm0.00002$ \citep{Ahumada2020}. The host galaxy is a star-forming galaxy with a low stellar mass of $\text{log}(M_{\star}/M_{\odot})=9.7$ and star formation rate (SFR) of 3.1 $M_{\odot}\ \text{yr}^{-1}$ \citep{Salim2016}. J1332$+$2634 is classified as a Seyfert 2 AGN \citep{Toba2014} and has a flux of 21.1 mJy at 1.4 GHz \citep{Condon1998}. The corresponding radio power at 1.4 GHz is $\text{log}(P_{\text{1.4 GHz}}/\text{W Hz}^{-1})=23.04$. 

The \hi\ absorption line profile of J1332$+$2634 is shown in Figure \ref{fig:abs_profile}, with the velocity Doppler corrected based on the optical redshift of the host galaxy. The detected \hi\ absorption line presents a multi-peak profile, implying unsettled kinematics of the absorbing gas. Fitted with a multicomponent Gaussian function, the line profile is decomposed into three velocity components centered at $-$72.0 $\text{km s}^{-1}$, $-$20.5 $\text{km s}^{-1}$, and 10.0 $\text{km s}^{-1}$, respectively. Despite the shallower blueshifted wing component, the velocity of the absorption system is in good agreement with the optical redshift of the host galaxy, suggesting the dominant components are likely tracing the circumnuclear rotating disk \citep{Gallimore1999}. The blueshifted wing centered at $-$72.0 $\text{km s}^{-1}$ has a narrow profile with an FWHM of 13.5 $\text{km s}^{-1}$. Adopting a 21.1 mJy continuum flux density at 1.4 GHz \citep{Condon1998}, we calculate the peak optical depth $\tau_{\text{peak}}=0.091\pm0.017$ and the integrated optical depth $\tau_{\text{int}}=4.91\pm0.67$ $\text{km s}^{-1}$. Based on Equation (\ref{eq:1}), the column density of this absorption system is $\sim 8.95\times10^{20}\ \text{cm}^{-2}$. 

\subsection{J1512\sl{+}3050} \label{subsec:J1512+3050}

The source J1512$+$3050 is an isolated galaxy with the spectroscopic redshift of $ z=0.08963\pm0.00002$ \citep{Ahumada2020}. The host galaxy is an elliptical galaxy with a high stellar mass of $\text{log}(M_{\star}/M_{\odot})=11.7$ \citep{Salim2016}. Classified as a LINER-type AGN \citep{Toba2014}, J1512$+$3050 has a 1.4 GHz continuum flux of 27.8 mJy \citep{Condon1998}, corresponding to a radio power at 1.4 GHz of $\text{log}(P_{\text{1.4 GHz}}/\text{W Hz}^{-1})=23.74$. 

As shown in Figure \ref{fig:abs_profile}, the \hi\ absorption line profile of J1512$+$3050 is Doppler corrected based on the optical redshift of the host galaxy. The multiple-peak profile of the \hi\ absorption line suggests the \hi\ gas is kinematically unsettled. A multicomponent Gaussian fit yields the peak optical depth $\tau_{\text{peak}}=0.107\pm0.019$ and the integrated optical depth $\tau_{\text{int}}=8.81\pm0.81$ $\text{km s}^{-1}$. The column density $\nhi$ of this absorption system based on Equation (\ref{eq:1}) is $\sim1.61\times10^{21}\ \text{cm}^{-2}$. The line profile is decomposed into one broad component centered at 74.9 $\text{km s}^{-1}$, and three narrow Gaussian components centered at 49.7 $\text{km s}^{-1}$, 120.2 $\text{km s}^{-1}$, and 167.5 $\text{km s}^{-1}$, respectively. The fitting parameters of each Gaussian component are listed in Table \ref{tab:fit}. The redshifted wings may suggest the \hi\ gas is clumpy and falling toward the AGN.

\subsection{J1628\sl{+}2529} \label{subsec:J1628+2529}

The source J1628$+$2529 is a member galaxy in a galaxy group with a spectroscopic redshift of $z=0.04010\pm0.00001$ \citep{Ahumada2020}. The host galaxy is an elliptical galaxy with a high stellar mass of $\text{log}(M_{\star}/M_{\odot})=11.1$ \citep{Salim2016}. Classified as a LINER-type AGN \citep{Jeong2013}, J1628$+$2529 has a 1.4 GHz continuum flux of 25.2 mJy \citep{Condon1998}, corresponding to a radio power at 1.4 GHz of $\text{log}(P_{\text{1.4 GHz}}/\text{W Hz}^{-1})=22.98$.

As shown in Figure \ref{fig:abs_profile}, the \hi\ absorption line profile of J1628$+$2529 is Doppler corrected and rebinned to a velocity resolution of 20 $\text{km s}^{-1}$, and the zero-point is corrected to the optical redshift of the host galaxy. Since this \hi\ absorption line shows a single-peak profile, we apply a single-component Gaussian fit to measure the peak optical depth $\tau_{\text{peak}}=0.072\pm0.017$ and the integrated optical depth $\tau_{\text{int}}=6.46\pm0.97$ $\text{km s}^{-1}$. Based on Equation (\ref{eq:1}), the column density $\nhi$ of this absorption system is $\sim1.18\times10^{21}\ \text{cm}^{-2}$. The line centroid is blueshifted $\sim -196\ \text{km s}^{-1}$ with respect to the systemic velocity of the host galaxy. Although the line width of this absorber is relatively narrow with a blueshifted centroid, it is likely the associated absorption of the AGN, considering the high column density. Since the velocity of the absorption line is significantly blueshifted, the \hi\ absorption with high column density may suggest the case of gas outflowing from the AGN.

\begin{deluxetable}{lcccccr}
\setlength\tabcolsep{1.5pt}
\tablenum{3}
\tablecaption{Gaussian fit parameters of the multi-peak \hi\ absorbers \label{tab:fit}}
\tabletypesize{\footnotesize}
\tablewidth{\textwidth}
\tablehead{
\colhead{Source name} & \colhead{FWHM} & \colhead{$v_{\text{centroid}}$} & \colhead{$\rm S/N_{peak}$} & \colhead{$\tau_{\text{peak}}$} & \colhead{$\tau_{\text{int}}$} & \colhead{$\nhi$}\\
\colhead{} & \colhead{($\text{km s}^{-1}$)} & \colhead{($\text{km s}^{-1}$)} & \colhead{} & \colhead{} & \colhead{($\text{km s}^{-1}$)} & \colhead{($10^{20} \text{cm}^{-2}$)}
}
\decimalcolnumbers
\startdata
J1332$+$2634 & 22.5 & 10.0 & 5.9 & 0.075 & 1.79 $\pm$ 0.47 & 3.26 $\pm$ 0.85 \\
 & 21.0 & -20.5 & 5.9 & 0.085 & 1.89 $\pm$ 0.47 & 3.44 $\pm$ 0.85 \\
 & 13.5 & -72.0 & 3.0 & 0.046 & 0.67 $\pm$ 0.35 & 1.22 $\pm$ 0.63 \\
J1512$+$3050 & 21.1 & 49.7 & 7.1 & 0.091 & 2.32 $\pm$ 0.58 & 4.23 $\pm$ 1.06 \\
 & 153.6 & 74.9 & - & 0.047 & 7.68 $\pm$ 1.36 & 14.00 $\pm$ 2.48 \\
 & 6.9 & 120.2 & 4.2 & 0.046 & 0.72 $\pm$ 0.41 & 1.31 $\pm$ 0.74 \\
 & 17.7 & 167.5 & 3.0 & 0.041 & 0.81 $\pm$ 0.57 & 1.48 $\pm$ 1.04
\enddata
\tablecomments{The columns are (1) the source name of the AGN; (2) full width at half maximum; (3) line centroid of the absorption component; (4) the S/N of identified peak in the spectrum; (5) peak optical depth; (6) integrated optical depth; (7) estimated \hi\ column density assuming $T_s=100$ K, $c_f=1$.}

\end{deluxetable}

\section{Discussions}\label{sec:dis}

\subsection{Detection Rate}\label{sec:det}
The pilot observations of \hi\ absorption lines suggest a detection rate of $\sim$$11.5^{+9.2}_{-3.7}\%$ in the sample of 26 low-power radio sources, and sources with $\text{log}(P_{\text{1.4 GHz}}/\text{W Hz}^{-1})<23$ present a lower detection rate of $\sim$$6.7^{+12.5}_{-2.2}\%$. Considering the uncertainties, the detection rate of \hi\  absorption in lower-power sources is relatively lower than that of previous targeted surveys through direct comparisons \citep{vanG1989,Emonts2010,Chandola2011,Ger2015,Maccagni2017}. However, a fair comparison of detection rates between surveys requires comparable detection limits and considerations of selection bias.

Since the representative sample observed by WSRT \citep{Ger2014,Ger2015,Maccagni2017} has overlapped regions with our sample in the radio power-redshift space (see Figure \ref{fig:dist}), the comparison between our sample and the WSRT sample under similar detection limits is reasonable and helpful to investigate the detection rate of low-power sources. In Figure \ref{fig:compare}(C), we presented the upper limits of peak optical depth ($\tau_{\text{peak}}$) for non-detections in both our sample and the WSRT sample \citep{Maccagni2017}. The upper limits of $\tau_{\text{peak}}$ were estimated with 3$\sigma$ rms noise (Table \ref{tab:a1}) at the velocity resolution of $\sim$16 $\text{km s}^{-1}$ based on Equation (\ref{eq:2}). Our observations achieved a median $\tau_{\text{peak}}$ upper limit of $\sim$0.04, which is similar to that of the WSRT observations \citep[$\sim$0.04,][]{Maccagni2017}. Thus, the detection limit of FAST observations is similar to that of \cite{Maccagni2017}.

Apart from the detection limit, selection effects can also significantly impact the detection rate of \hi\ absorption lines. Previous studies have revealed that a large fraction of early-type galaxies harbors amounts of detectable \hi\ gas \citep{Oosterloo2010,Serra2012}. Therefore, the elimination of gas-rich sources in the AGN sample will decrease the detection probability of \hi\ absorption and reduce a large fraction of the parent population. In the selection of our parent sample, we excluded sources that already have \hi\ emission detections (S/N$>5$) to reduce redundant observations, because FAST observations focus on searching \hi\ absorption lines to build up an \hi\ absorption sample of faint radio AGNs. After crossmatching with HI-MaNGA DR2 and ALFALFA catalogs, we excluded 19 out of 180 sources in the parent sample, corresponding to approximately two to three potentially excluded sources in the pilot observations. The rest sources of the parent sample were not observed by either HI-MaNGA or ALFALFA, and the sample still contains a certain amount of gas-rich ($\sim$35\%) galaxies based on SDSS spectra. The potential exclusion of two to three \hi-rich sources could reduce the chance of detecting new \hi\ absorption lines in the sample, which may underestimate the detection rate. Therefore, the \hi\ absorption detection rate of $11.5^{+9.2}_{-3.7}\%$ can be considered as the lower limit for our sample. For sources with $\text{log}(P_{\text{1.4 GHz}}/\text{W Hz}^{-1})<23$, the \hi\ absorption detection rate of $6.7^{+12.5}_{-2.2}\%$ could also represent the lower limit.


For our sample observed by FAST, the \hi\ emission and the radio morphology of sources may reduce the likelihood of detecting \hi\ absorption. Since the \hi\ emission becomes more detectable at a lower redshift, the lower detection rate of \hi\ absorption may be partly due to the dilution by \hi\ emission \citep{Allison2014,Curran2018}. In the selection of the parent sample, most of the excluded gas-rich sources distribute at lower redshifts ($z<0.03$), and their potential \hi\ absorptions could be diluted by \hi\ emission. In addition, previous studies have found the detection rate of \hi\ absorption in compact sources is higher than that in extended sources \citep[e.g.,][]{Maccagni2017}. We followed \cite{Ger2015} and \cite{Maccagni2017} to inspect our sources, and the classification results suggest our sample has a large fraction of extended radio sources ($\sim$65\%, Table \ref{tab:a1}), which may also lead to the observed lower detection rate. Our sample has nine compact sources ($\sim$35\%) with two detected absorptions, and one detection out of 17 extended sources ($\sim$65\%). The detection rate of the compact sources in our sample ($22.2^{+18.3}_{-8.1}$\%) is consistent with that of Maccagni et al. (2017) within error bars. Considering the uncertainties, the detection rate of the extended sources in our sample ($5.9^{+11.3}_{-1.9}$\%) is consistent with that of Maccagni et al. (2017). Therefore, the lower detection rate of \hi\ absorption in FAST observations may be due to the dilution by \hi\ emission at lower redshift and the selection of more extended sources.

Thus, as a reasonable comparison of detection rates between our sample and the WSRT sample, our observations suggest a lower detection rate of \hi\ absorption lines in the low-power radio sources. Due to the incompleteness of the sample, these detection rates may represent the lower limits. The selection of more extended sources and dilution by \hi\ emission at lower redshift may contribute to the lower detection rate of \hi\ absorption lines. One caveat of this result is the fact that the detection rate could be affected by low number statistics, which will be further investigated with the complete survey of the parent sample.

\subsection{Kinematics and the Origin of the Absorbing Gas}\label{sec:kin}

To investigate the origin of \hi\ 21 cm absorption, many studies have used high spatial resolution interferometry observations to explore the detailed kinematics and structures of the absorption. These observations have found evidence of various probabilities that can form \hi\ absorption in radio AGNs: rotating gas disks \citep{Gallimore1999,Morganti2008,Struve2010}, high-velocity clouds \citep[HVCs;][]{Conway1999,Struve2012}, gas outflows \citep{Morganti2013,Schulz2021}, and gas falling toward the supermassive black hole \citep{Morganti2009,Maccagni2014,Tremblay2016}. Based on these observations, the origin and related mechanisms of the \hi\ absorption can be discussed by analyzing the line profiles and kinematics of the absorbing gas. Alternatively, since the absorption line traces the foreground gas along the line of sight of the continuum source, the morphology and structure of the radio continuum can also contribute to the observed absorption profiles \citep{Murthy2021}.

\begin{figure*}[t!]
\includegraphics[width=\textwidth]{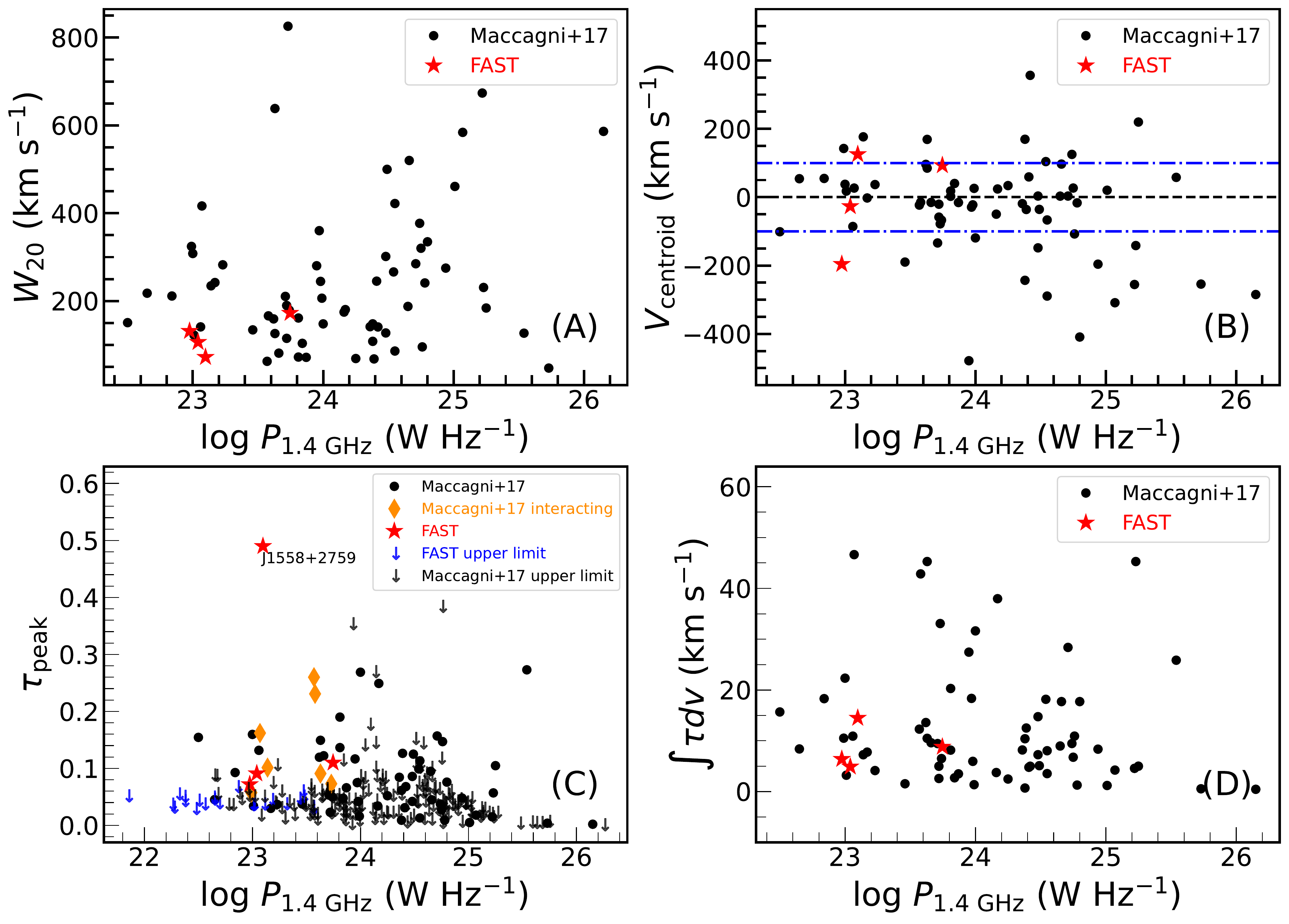}
\caption{(A) Line width measured at 20\% of the intensity ($W_{20}$) of the \hi\ profiles vs. the radio power of the sources. (B) Line centroid offset with respect to the systemic velocity vs. the radio power of the sources. The fine dashed lines in blue show the interval of $\pm 100\ \text{km s}^{-1}$. (C) Peak optical depth vs. the radio power of the sources. The interacting sources of \cite{Maccagni2017} are marked with orange diamonds. (D) Integrated optical depth vs. the radio power of the sources.} 
\label{fig:compare}
\end{figure*}

Figure \ref{fig:compare}(A) shows the $W_{20}$ of the \hi\ absorption lines versus the radio power of the sources. \cite{Maccagni2017} has found that broad lines ($W_{20}>400\ \text{km s}^{-1}$) were only detected in the interacting galaxies at low radio power ($\text{log}(P_{\text{1.4 GHz}}/\text{W Hz}^{-1})<24$). The line widths of the new \hi\ absorbers detected by FAST are relatively narrow and consistent with the results of \cite{Maccagni2017}. One of our sources is discovered in a merging galaxy pair \citep{Yu2022}, and comparable narrow lines also have been detected in the interacting sources \citep{Maccagni2017,Dutta2019}. In Figure \ref{fig:compare}(B), we plot the centroid shift of the line with respect to the galaxy's systemic velocity versus the radio power of the sources. The fine dashed lines in blue show the interval of $\pm 100\ \text{km s}^{-1}$. \cite{Maccagni2017} suggests the majority of the lines detected in low-power sources are centered at the systemic velocity ($|\Delta v|< 100\ \text{km s}^{-1}$), while two of four in our FAST detected sources are offset relative to the systemic velocity. For the source J1512$+$3050, although the systemic centroid of the absorbing gas is $\sim$93 $\text{km s}^{-1}$, it still has two narrow components redshifted more than 100 $\text{km s}^{-1}$. Our detections have a high fraction of sources with large velocity offset ($|\Delta v|> 100\ \text{km s}^{-1}$), which may suggest various kinematics and possible connections to the inflows and outflows of the gas.

Previous observations suggest the \hi\ 21cm absorption originated from rotating disks generally have symmetric profiles and line widths of a few hundred $\text{km s}^{-1}$, centered at or close to the systemic velocity \citep{Gallimore1999,Murthy2021}. The absorption lines associated with gas outflows usually show broad and asymmetric profiles, and the line centroid is blueshifted from the systemic velocity \citep{Morganti2013,Schulz2021}. Relatively narrow redshifted profiles may originate from gas clouds falling into the SMBH \citep{Morganti2009,Maccagni2014}. Alternatively, HVCs along the line of sight can also be observed as narrow \hi\ absorption lines \citep{Conway1999,Araya2010}. Besides, the structures (e.g., core, jets, or lobes) of the background continuum source may partially align with the \hi\ gas, which can result in the multiple velocity components and asymmetries observed in the absorption profiles \citep{Peck2001,Struve2010b,Murthy2021}. In particular, for the radio AGN with the typical two-jetted structure, since one jet is often behind the \hi\ gas, the jet-cloud alignment can produce asymmetric or velocity-shifted absorption profiles even if the \hi\ disk and radio continuum both have symmetric morphologies.

The \hi\ absorption profile of J1332$+$2634 shows a symmetric shape of the main component centered at the systemic velocity of the host galaxy, which indicates that a rotating disk very likely contributes the majority of absorption. However, J1332$+$2634 presents a shallower narrow wing centered at $-$72.0 $\text{km s}^{-1}$, with a column density of $\sim$$1.22\times10^{20}\ \text{cm}^{-2}$. Since confirmed \hi\ outflows generally present broad profiles with large blueshifted velocity, the narrow wing of absorption in J1332$+$2634 likely traces an intermediate velocity cloud (IVC). Additionally, with spatially unresolved observations, we cannot rule out the scenario that the observed wing component or asymmetric absorption profile may be due to structures of the radio continuum, so the foreground \hi\ gas could distribute in a regular rotating disk.

The absorption profile of J1512$+$3050 presents an asymmetric multi-peak shape, redshifted with respect to the systemic velocity of the host galaxy. These two shallower redshifted components show narrow profiles with column densities of $\sim$$1.31\times10^{20}\ \text{cm}^{-2}$ and $\sim$$1.48\times10^{20}\ \text{cm}^{-2}$, respectively. The redshifted wings may suggest gas clouds falling onto the SMBH, which is correlated to the feeding and triggering of AGN. Alternatively, clumpy HVCs along the line of sight may also produce the presented absorption profile. Considering the possible partial alignment between the continuum structures (e.g., lobes) and the \hi\ cloud, the warped \hi\ could also present such features \citep{Murthy2021}.

The absorption profile of J1628$+$2529 shows a narrow blueshifted shape, and the offset relative to systemic velocity is $-$195.9 $\text{km s}^{-1}$. Although the line width of the absorption is relatively narrow compared to the broad lines revealed by previous observations \citep[e.g.,][]{Morganti2013,Schulz2021}, gas outflows from the AGN could be the possible origin. In particular, observations at higher redshift have discovered a narrow ($\sim$40 $\text{km s}^{-1}$) blueshifted \hi\ line that traced the outflows driven by the radio jet \citep{Aditya2017}. As shown in Figure \ref{fig:compare}(B), low-power radio sources ($\text{log}(P_{\text{1.4 GHz}}/\text{W Hz}^{-1})<24$) with large blueshifted velocity offsets are relatively rare compared to high-power radio sources ($\text{log}(P_{\text{1.4 GHz}}/\text{W Hz}^{-1})>24$). Especially, for sources with $\text{log}(P_{\text{1.4 GHz}}/\text{W Hz}^{-1})<23$, J1628$+$2529 is the only detection of sources with large blueshifted velocity offsets ($\Delta v< -100\ \text{km s}^{-1}$). If the \hi\ absorption line of J1628$+$2529 is confirmed to be outflows, it could be the lowest radio power source with outflows. Assuming the absorbing gas to be outflows, we can estimate the mass outflow rate following \cite{Heckman2002}:
\begin{equation}
    \dot{M}_{\hi}\sim 30\frac{r_{\star}}{\rm kpc}\frac{\nhi}{\rm 10^{21}\ cm^{-2}}\frac{v}{\rm 300\ km\ s^{-1}}\frac{\Omega}{4\pi}\ M_{\odot}\ {\rm yr^{-1}},
\end{equation}
where the $r_{\star}$ is the radius of the outflow gas, $\nhi$ is the estimated column density assuming $T_s=100$ K, $c_f=1$, $v$ is the velocity of the outflow gas, and $\Omega$ is the solid angle of the gas assumed to be $\pi$. Since neither FAST nor VLA resolves the compact continuum source, we assume a typical upper limit radius of $r_{\star}=1$ kpc considering outflows generally appear to be limited to the central kiloparsec scale \citep{Morganti2018}. Therefore, we estimate the mass outflow rate to be $\sim$5.8 $M_{\odot}\ {\rm yr^{-1}}$, with $\nhi=1.18\times10^{21}\ \text{cm}^{-2}$ and $r_{\star}=1$ kpc. Alternatively, HVCs away from the nucleus could also result in the blueshifted absorption feature. Besides, for a typical two-jetted radio AGN, the jet-cloud alignment may give rise to the significantly blueshifted profile when one jet is behind the rotating \hi\ disk.

Similarly, the absorption profile of J1558$+$2759 is narrow and redshifted, which can be contributed either by gas fueling or galactic gas clouds. Observations of \hi\ absorptions in mergers have detected significant systemic velocity redshift, which suggests the signature of gas infall \citep{Dutta2019}. Since J1558$+$2759 is an interacting source, gas driven by the merging process could infall toward the AGN, which may connect with the gas fueling onto the SMBH \citep{Srianand2015}. On the other hand, HVCs or a large-scale gas disk along the line of sight can also be observed as redshifted absorption lines.

Since FAST cannot resolve these sources, spatially resolved \hi\ observations by interferometry will help understand the kinematics and origin of the absorbing gas, as well as possible jet-ISM interactions.

\subsection{Uncertainties of the Systemic Velocity}\label{sec:uncertain}
To understand the origin of the absorbing gas, the important information of gas kinematics depends on the comparison between the systemic velocity of the galaxy and the velocity of \hi\ gas. As revealed by previous studies \citep[e.g.,][]{Morganti2001}, the uncertainties of the redshift of the galaxy could bias the derived systemic velocity and further affect our understanding of the origin of the gas.

For our detected sources, the systemic velocities are derived by optical redshift from SDSS spectroscopy. The 1$\sigma$ error of the redshift is $\sim 0.00002$, which corresponds to 6 $\text{km s}^{-1}$. Considering the spectral resolution of the profile, the line centroid of J1332$+$2634 is $-26.8\pm11.7$ $\text{km s}^{-1}$, which indicates the velocity offset of \hi\ absorption line is not significant with respect to the systemic velocity. The source J1512$+$3050 has a velocity offset of $93.0\pm11.7$ $\text{km s}^{-1}$, which suggests the \hi\ absorption line is significantly redshifted as discussed above. The line centroid of J1628$+$2529 is $-195.9\pm20.9$ $\text{km s}^{-1}$, indicating the \hi\ absorbing gas is significantly blueshifted relative to the systemic velocity. For the interacting source J1558$+$2759, an older study claimed a spectroscopic redshift of $ z=0.05174\pm0.00027$ \citep{Kim1995}, which corresponds to a larger error of 81 $\text{km s}^{-1}$. If the older redshift is adopted to derive the systemic velocity, the velocity offset of the \hi\ absorption line becomes $23.4\pm81.0$ $\text{km s}^{-1}$. The old observations with large uncertainties could be limited by the previous instruments, while the SDSS dataset contains spatially resolved MaNGA observations, which can alleviate uncertainties from motions of emitting gas. Thus, the redshift from SDSS is likely more accurate to derive the real systemic velocity, and the source J1558$+$2759 is significantly redshifted with respect to the systemic velocity.  

\subsection{Optical Depth}
In Figure \ref{fig:compare}(C) and (D), we show a comparison of the peak optical depth and the integrated optical depth of our sources with that of \cite{Maccagni2017}. The peak optical depths of our sources are consistent with previous observations, except for the interacting source J1558$+$2759, which presents remarkably high $\tau_{\text{peak}}$. The high $\tau_{\text{peak}}$ value of J1558$+$2759 is close to that of \hi\ absorbers detected at higher redshift \citep{Chowdhury2020}. We have marked the interacting sources of \cite{Maccagni2017} with orange to check whether all the interacting sources show higher $\tau_{\text{peak}}$ values. The mean $\tau_{\text{peak}}$ value of the interacting sources ($0.14\pm0.03$) is slightly higher than that of noninteracting sources ($0.08\pm0.01$). In addition, a recent survey of \hi\ absorption lines in the galaxy merger sample \citep{Dutta2018,Dutta2019} also reveals a few sources with higher peak optical depth ($\tau_{\text{peak}}>0.4$). This provides a hint that galaxy interactions and mergers may have contributions to the high value of $\tau_{\text{peak}}$. Observations and simulations suggest galaxy interactions can induce tidal torque to trigger the infall of diffuse ionized halo gas toward the central region of the galaxy, which increases the density of the ionized gas and makes the cooling more efficient \citep{Braine1993,Moreno2019}. In this scenario, the ionized gas will turn into atomic gas, and the external pressure (e.g., shocks) induced by interactions will accelerate the transition from atomic to molecular gas \citep{Barnes2004,Kaneko2017}, which would create a dense environment. Indeed, previous observations of \hi\  absorption lines have detected significant high $\nhi$ in mergers \citep{Dutta2019}, which may result in the high $\tau_{\text{peak}}$ of the interacting sources. For the integrated optical depth, the results of our detections are consistent with those of \cite{Maccagni2017}.

\section{Summary}\label{sec:sum}

We present pilot observations of \hi\ absorption search toward low-power radio AGNs by FAST. With the high sensitivity of FAST, we discovered three new \hi\ absorbers at a very weak flux level. Our pilot observations suggest the detection rate of \hi\ absorption lines in our sample is $\sim$$ 11.5^{+9.2}_{-3.7}\%$. Considering the uncertainties, the detection rate is slightly lower than that of brighter sources, but the difference is not significant ($<2\sigma$). Our results indicate that the low-power sources with $\text{log}(P_{\text{1.4 GHz}}/\text{W Hz}^{-1})<23$ may have a lower detection rate of \hi\ absorption, and the detection rate increases and remains similar as the radio power elevating. Due to the incompleteness of the sample, these detection rates may represent the lower limits. The selection of more extended sources and dilution by \hi\ emission at lower redshift may contribute to the lower detection rate of \hi\ absorption lines. These detected absorbers present relatively narrow line widths and comparable column densities as revealed by previous observations of low-power radio AGNs. One absorber has a symmetric profile with a large velocity offset, while the other two absorbers show asymmetric profiles that can be decomposed into multiple components, suggesting various possibilities of gas kinematics and origins. These absorbers may have connections with rotating disks, gas outflows, galactic gas clouds, gas fueling of the AGN, and jet-ISM interactions, which needs to be confirmed or ruled out by spatially resolved observations. The upcoming systematic survey and high spatial resolution observations of the \hi\ absorbers in low-power AGNs will shed new light on our understanding of multiple evolution processes of the faint radio AGN population. 

\begin{acknowledgments}
We thank the anonymous referee for the careful reading and helpful comments that improved the paper. Q.Y. appreciates the helpful suggestions by Dr. Mouyuan Sun, and observing support from Dr. Bo Zhang. This work is supported by the National Key R\&D Program of China under No. 2017YFA0402600, and the National Natural Science Foundation of China under Nos. 11890692, 12133008, 12221003. Junfeng Wang acknowledges support by the NSFC grants Nos. U1831205 and 12033004. We acknowledge the science research grants from the China Manned Space Project with Nos. CMS-CSST-2021-A04, CMS-CSST-2021-A05, CMS-CSST-2021-A06, CMS-CSST-2021-B02.

This work made use of the data from FAST (Five-hundred-meter Aperture Spherical radio Telescope).  FAST is a Chinese national mega-science facility, operated by National Astronomical Observatories, Chinese Academy of Sciences. 
\end{acknowledgments}

%

\vspace{5mm}
\facilities{FAST}


\software{Astropy \citep{Astropy2013},  
          Scipy \citep{SciPy2020}, 
          LMFIT \citep{lmfit2014}.
          }


\appendix
\section{Summary Table of Non-detections}
We summarized the ancillary information of the \hi\ absorption non-detections in Table \ref{tab:a1}.

\begin{deluxetable*}{lccccccccc}
\tablenum{A1}
\tablecaption{ \hi\ Absorption Non-detections\label{tab:a1}}
\tabletypesize{\small}
\tablewidth{\textwidth}
\tablehead{
\colhead{Source name} & \colhead{R.A.} & \colhead{Decl.} &
\colhead{$t_{\text{ON}}$} & \colhead{$z$} & \colhead{$S_{\text{1.4\ GHz}}$} & \colhead{log($P_{\text{1.4\ GHz}}$)} & \colhead{rms} & \colhead{$\tau_{\text{peak}}$} & \colhead{Radio type}\\
\colhead{} & \colhead{deg.} & \colhead{deg.} &
\colhead{(min)} & \colhead{} & \colhead{(mJy)} & \colhead{(W $\text{Hz}^{-1}$)} & \colhead{(mJy)} & \colhead{} & \colhead{}
}
\decimalcolnumbers
\startdata
J0312$-$0004 & 48.01039 & $-$0.07847 & 12.3 & 0.03762 & 14.7 & 22.68 & 0.19 & $<$0.040 & E\\
J0322$-$0000 & 50.63664 & $-$0.00121 & 12.3 & 0.02175 & 17.2 & 22.27 & 0.18 & $<$0.032 & E\\
J0352$-$0620 & 58.05244 & $-$6.34894 & 10.7 & 0.03295 & 12.5 & 22.50 & 0.22 & $<$0.044 & E\\
J0406$-$0506 & 61.68081 & $-$5.11472 & 12.3 & 0.06600 & 14.2 & 23.18 & 0.21 & $<$0.045 & E\\
J0735$+$4212 & 113.98199 & 42.20336 & 12.3 & 0.08822 & 18.5 & 23.56 & 0.20 & $<$0.033 & C\\
J0746$+$3029 & 116.63530 & 30.49075 & 12.3 & 0.05704 & 16.3 & 23.10 & 0.20 & $<$0.038 & E\\
J0839$+$2308 & 129.96150 & 23.14335 & 21.3 & 0.02520 & 12.3 & 22.25 & 0.15 & $<$0.037 & C\\
J0850$+$5522 & 132.59502 & 55.37874 & 5.0 & 0.03076 & 20.0 & 22.64 & 0.30 & $<$0.046 & E\\
J0903$+$4026 & 135.75897 & 40.43399 & 12.3 & 0.02877 & 15.5 & 22.47 & 0.18 & $<$0.029 & C\\
J0920$+$0102 & 140.00904 & 1.03830 & 21.3 & 0.01703 & 10.7 & 21.84 & 0.17 & $<$0.052 & E\\
J1022$+$3634 & 155.58076 & 36.58304 & 5.3 & 0.02592 & 13.4 & 22.31 & 0.24 & $<$0.055 & E\\
J1106$+$4602 & 166.65570 & 46.03878 & 6.2 & 0.02501 & 16.3 & 22.37 & 0.23 & $<$0.043 & E\\
J1120$+$5049 & 170.16647 & 50.82729 & 5.0 & 0.02762 & 24.8 & 22.64 & 0.40 & $<$0.050 & E\\
J1124$+$4708 & 171.24463 & 47.14351 & 6.2 & 0.05372 & 10.5 & 22.86 & 0.23 & $<$0.068 & C\\
J1140$+$4632 & 175.19730 & 46.54049 & 12.3 & 0.05359 & 14.9 & 23.01 & 0.18 & $<$0.037 & E\\
J1153$+$5241 & 178.37742 & 52.68945 & 2.5 & 0.07164 & 23.0 & 23.46 & 0.45 & $<$0.060 & E\\
J1340$+$2629 & 205.23577 & 26.48672 & 5.0 & 0.07502 & 26.0 & 23.55 & 0.25 & $<$0.029 & C\\
J1423$+$4015 & 215.96472 & 40.25883 & 21.3 & 0.08220 & 12.1 & 23.31 & 0.15 & $<$0.038 & E\\
J1445$+$5134 & 221.43798 & 51.58082 & 16.0 & 0.02963 & 11.5 & 22.37 & 0.19 & $<$0.051 & C\\
J1539$+$4438 & 234.87364 & 44.64845 & 6.5 & 0.07299 & 20.6 & 23.43 & 0.32 & $<$0.048 & C\\
J1544$+$4330 & 236.21348 & 43.51406 & 21.3 & 0.03664 & 11.4 & 22.55 & 0.14 & $<$0.038 & E\\
J1624$+$2507 & 246.09814 & 25.13012 & 16.0 & 0.09603 & 12.6 & 23.47 & 0.21 & $<$0.051 & E\\
J2128$+$0017 & 322.04858 & 0.29988 & 21.3 & 0.05250 & 15.2 & 23.00 & 0.18 & $<$0.036 & E
\enddata
\tablecomments{The columns are (1) the source name of the AGN; (2) R.A. in degrees; (3) Decl. in degrees; (4) integration time; (5) redshift from SDSS spectroscopy \citep{Abolfathi2018}; (6) radio continuum flux at 1.4 GHz from NVSS \citep{Condon1998}; (7) radio power at 1.4 GHz; (8) 1$\sigma$ noise measured at the velocity resolution of 16 $\text{km s}^{-1}$; (9) upper limit of the peak optical depth estimated with 3$\sigma$ rms; (10) radio morphology classification as compact (C) and extended (E). }

\end{deluxetable*}

\bibliography{ref}{}
\bibliographystyle{aasjournal}


\end{CJK*}
\end{document}